\documentclass[10pt,journal,compsoc]{IEEEtran}
\usepackage{booktabs} 
\usepackage{bm}
\usepackage{color}
\usepackage{subfigure}
\usepackage{multirow}
\usepackage{graphicx}
\usepackage{amssymb}

\def\L{\mathcal{L}}

\def\v{\bm{v}} 
\def\h{\bm{h}} 
 
\def\n{\bm{n}}

\def\x{\bm{x}} 
\def\y{\bm{y}} 
\def\modela{GFM} 
\def\modelb{CD-GFM}
\def\setR{\mathbb{R}}
\def\setU{\mathbb{U}}
\def\setI{\mathbb{I}}
\def\D{\mathcal{D}} 
\def\N{\mathcal{N}}
\def\Y{\mathcal{Y}}

\ifCLASSOPTIONcompsoc
  \usepackage[nocompress]{cite}
\else
  \usepackage{cite}
\fi
\ifCLASSINFOpdf
\else
\fi
\begin{document}
%
\title{Graph Factorization Machines for Cross-Domain Recommendation}
%
%
%
%

\author{Dongbo Xi, Fuzhen Zhuang, Yongchun Zhu, Pengpeng Zhao, Xiangliang Zhang and Qing He
\IEEEcompsocitemizethanks{\IEEEcompsocthanksitem Dongbo Xi, Fuzhen Zhuang, Yongchun Zhu and Qing He are with the Key Laboratory of Intelligent Information Processing of Chinese Academy of Sciences (CAS), Institute of Computing Technology, CAS, Beijing 100190, China and the University of Chinese Academy of Sciences, Beijing 100049, China. \protect \\ E-mail: \{xidongbo17s,zhuangfuzhen,zhuyongchun18s,heqing\}@ict.ac.cn
\IEEEcompsocthanksitem Pengpeng Zhao is with Soochow University.
\protect \\ E-mail: ppzhao@suda.edu.cn
\IEEEcompsocthanksitem Xiangliang Zhang is with the Machine Intelligence and Knowledge Engineering (MINE) Laboratory at the Division of Computer, Electrical, and Mathematical Sciences and Engineering, King Abdullah University of Science and Technology (KAUST), Saudi Arabia.
\protect \\ E-mail: xiangliang.zhang@kaust.edu.sa}
}

%
%

\markboth{}%
{Shell \MakeLowercase{\textit{et al.}}: Bare Demo of IEEEtran.cls for Computer Society Journals}
%



\IEEEtitleabstractindextext{%
\begin{abstract}
Recently, graph neural networks (GNNs) have been successfully applied to recommender systems. In recommender systems, the user's feedback behavior on an item is usually the result of multiple factors acting at the same time. However, a long-standing challenge is how to effectively aggregate multi-order interactions in GNN. In this paper, we propose a Graph Factorization Machine (\modela) which utilizes the popular Factorization Machine to aggregate multi-order interactions from neighborhood for recommendation.
Meanwhile, cross-domain recommendation has emerged as a viable method to solve the data sparsity problem in recommender systems. However, most existing cross-domain recommendation methods might fail when confronting the graph-structured data. In order to tackle the problem, we propose a general cross-domain recommendation framework which can be applied not only to the proposed \modela, but also to other GNN models.
We conduct experiments on four pairs of datasets to demonstrate the superior performance of the GFM. Besides, based on general cross-domain recommendation experiments, we also demonstrate that our cross-domain framework could not only contribute to the cross-domain recommendation task with the GFM, but also be universal and expandable for various existing GNN models.
\end{abstract}

\begin{IEEEkeywords}
Graph factorization machines, graph neural networks, factorization machines, cross-domain recommendation.
\end{IEEEkeywords}}

\maketitle

\IEEEdisplaynontitleabstractindextext

%
\IEEEpeerreviewmaketitle

\IEEEraisesectionheading{\section{Introduction}\label{sec:introduction}}
%
%
%
%
\IEEEPARstart{W}{ith} the explosively growing of personalized online applications,
recommender systems have been widely used in various real scenarios, such as recommending movies to watch at MovieLens and products to purchase at Amazon.
Data collected by these online applications have been effectively leveraged for studying users' online activities and patterns, which provides unparalleled opportunities to built personalized recommender systems.

Collaborative filtering is a widely adopted method \cite{koren2009matrix,he2017neural} in recommender systems, which assumes that users and items that are similar in history will also be similar in future. Recently, more and more studies have found that the graph-structured data are of great benefit to improve the performance of recommender systems~\cite{wang2019app3,li2019fignn}. However, it is hard for the traditional collaborative filtering methods to utilize the graph-structured data. In recent years, considerable efforts have been made to learn from the graph-structured data via the Graph Neural Network (GNN) \cite{kipf2016gcn,velivckovic2017gat,hamilton2017graphsage}, and we have also witnessed the rapid development and popularity of graph neural networks in recommender systems \cite{wang2019app3,li2019fignn}. The main intuition behind this line of methods is that the latent representation of a node could be integrated by iteratively transforming, propagating, and aggregating node features from its local neighborhood \cite{ding2019graph}. 

In the recommendation field, as is known to all, data sparsity is a major problem, and it is important to effectively exploit the sparse data~\cite{rendle2010factorization,he2017nfm,guo2017deepfm,lian2018xdeepfm} to capture multi-factor interaction information. However, the aggregating schemes of these GNN-based methods are too simplistic, e.g., the mean or pooling~\cite{hamilton2017graphsage}, which is \textbf{difficult to capture sufficient interaction information from neighborhood}. 

In order to tackle the above problem, we propose a novel \textbf{Graph Factorization Machine} (\modela) with the advantage of popular Factorization Machine (FM) \cite{rendle2010factorization}. FM has been successful to effectively exploit sparse data and capture feature interactions for recommender systems \cite{rendle2010factorization, he2017nfm,guo2017deepfm}, but it cannot work with graph-structure data. To this end, we utilize FM in our \modela~to aggregate the second-order neighbor messages, and the superposition of multiple \modela~layers to aggregate  the higher-order neighbor messages. 

Besides, to address the data sparsity in recommender systems, leveraging auxiliary data from other domains, cross-domain recommendation ~\cite{pan2010cst,tang2012ctl,man2017emcdr,chen2019eatnn,hu2019hybrid,yuan2019darec} is an effective method.
However, most existing cross-domain recommendation methods might fail when confronting the graph-structured data.
In order to tackle the problem, we propose a general cross-domain recommendation framework which can be applied not only to the proposed \modela~to form the cross-domain GFM (\modelb), but also to other GNN models, e.g., the GCN \cite{kipf2016gcn}, GAT \cite{velivckovic2017gat}, GraphSAGE \cite{hamilton2017graphsage} and so on.
On the one hand, the framework utilizes shared node representations to initialize the graph nodes in the source and target domains for learning domain-shared features, and these shared nodes can be users or items, or both, which no longer has to assume specific sharing patterns.
On the other hand, the framework uses the graph structure data of each domain to learn the domain-specific features and cooperates the learning of the graph topology through sharing the graph parameters.
Finally, domain-shared and domain-specific features are combined in each domain and used as prediction tasks. 

To summarize, the contributions of this paper are as follows:
\begin{itemize} 
\item We propose a novel GNN model called Graph Factorization Machine (\modela) to capture features more effectively with graph-structure data than existing GNN-based methods. 
\item We propose a general cross-domain recommendation framework, which can be naturally applied not only to the proposed \modela~to form the cross-domain GFM (\modelb), but also to other GNN models.
\item We perform experiments on four pairs real-world datasets to demonstrate the effectiveness of the \modela. In addition, we demonstrate our cross-domain recommendation framework is general for various existing GNN models on both user-shared and item-shared cross-domain tasks.
\end{itemize}

\section{Related Work}
In this section, we present the related work in three-folds: Factorization Machines, Graph Neural Networks and Cross-Domain Recommendation.

\subsection{Factorization Machines}
Many prediction tasks need to address categorical variables (e.g., user IDs, item IDs, etc) for obtaining excellent performance, a popular solution is to convert them to binary features via one-hot encoding, but the encoding is high-dimension and sparse.
Factorization Machine (FM) \cite{rendle2010factorization} is a widely used method to model second-order feature interactions automatically from such high-dimension and sparse one-hot features via the inner product of raw embedding vectors.
For combining the advantages of the FM on modeling the second-order and the neural network on modeling the higher-order feature interactions, some studies have extended the FM to neural networks \cite{zhang2016fnn,qu2016pnn,he2017nfm,guo2017deepfm}.
For example, Factorization-machine supported Neural Networks (FNN) \cite{zhang2016fnn}
utilizes the pre-trained factorization machine as the bottom layer of a multi-layer neural networks.
Product-based Neural Networks (PNN) \cite{qu2016pnn} utilizes an embedding
layer to learn a distributed representation of the categorical data, a product layer to capture interactive patterns between inter-field categories, and further fully connected layers to explore high-order feature interactions.
A method called Neural Factorization Machine (NFM) \cite{he2017nfm} has also been proposed to use a Hadamard product based FM followed by the MLP.
Guo et al. proposed a Factorization Machine based Neural Network (DeepFM) \cite{guo2017deepfm}, whose ``wide" (FM) and ``deep" (MLP) parts have a shared input and are fed to the output layer in parallel.
Some other approaches have also attempted to learn higher-order feature interactions explicitly instead of the implicit ``deep" part \cite{wang2017deepcross,lian2018xdeepfm}.
However, the above methods might fail when confronting the graph-structured data innate in recommender systems.

\subsection{Graph Neural Networks}
Graph neural networks (GNNs) are deep learning based methods that operate on graph-structured data. Due to its convincing performance and high interpretability, GNN has been a widely applied graph learning method and achieved remarkable performance \cite{kipf2016gcn,hamilton2017graphsage,velivckovic2017gat,ding2019graph,grover2016app1,wang2019app3} recently.
The concept of GNN was first proposed in \cite{scarselli2008graph}, which is the pioneer work to learn graph node representations based on neural networks. 
By designing different schemes for the graph convolutional layer, a lot of graph convolutional networks (GCNs) have emerged and demonstrated superior on learning node representations based on the graph spectral theory.
The simplified GCN \cite{kipf2016gcn} utilized a linear filter and achieved better performance.
The most of the prevailing GNN models followed the neighborhood aggregation strategy, and proposed  different schemes for message aggregation.
Among them, Graph Attention Networks (GATs) \cite{velivckovic2017gat} have been proposed to learn different weights via attention mechanism for the neighbor messages when aggregating neighbors.
GraphSAGE \cite{hamilton2017graphsage} designed mean/LSTM/pooling three different message aggregators to aggregate the neighbor messages.
However, the aggregating schemes of these GNN methods are too simplistic and are not suitable in recommender systems, in which multiple factor interactions are more efficient.

\subsection{Cross-Domain Recommendation}
Cross-domain recommendation can take the advantage of existing large scale data in the source domain and improve the data sparsity and recommendation quality in the related target domain.
Traditional methods such as the Coordinate System Transfer (CST) \cite{pan2010cst} aimed to discover the principle coordinates of both users and items in the source data matrices, and transfer them to the target domain in order to reduce the effect of data sparsity.
Some works \cite{tang2012ctl,loni2014cross,liu2015non} extended the classical Collaborative Filtering (CF) to the cross-domain scenario. 
Recently, neural networks have been used to implement cross-domain recommendation.
For example, Chen et al. proposed to introduce attention mechanisms to automatically assign a personalized transfer scheme for each user \cite{chen2019eatnn}.
There are various sharing scenarios in these cross-domain recommendation researches, such as the sharing users \cite{hu2013personalized,yan2019tciqi,yuan2019darec,hu2019hybrid}, sharing items \cite{gao2019natr}, sharing accounts \cite{ma2019pi} and so on.
However, most of these existing cross-domain recommendation methods were designed for traditional structured data.
They might fail when encountering massive graph data in recommender systems.

\section{Methodology}
\begin{figure*}[!t]
	\begin{center}
		\includegraphics[width=0.7\linewidth]{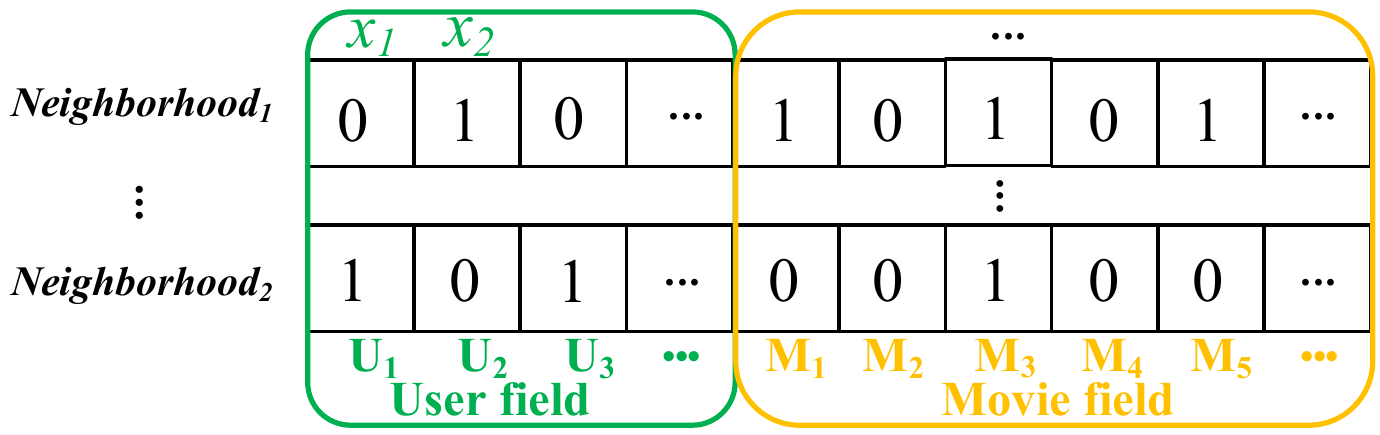}
		\vspace{-2mm}
		\caption{Two neighborhood examples in user-item interaction graph. In the $Neighborhood_1$, the user $U_2$ watched movies $M_1, M_3, M_5$, and in the $Neighborhood_2$, the movie $M_3$ was watched by the users $U_1, U_3$.}
		\label{fig:example}
	\end{center}
\end{figure*}
\begin{figure*}[!t]
\centering
\includegraphics[width=\linewidth]{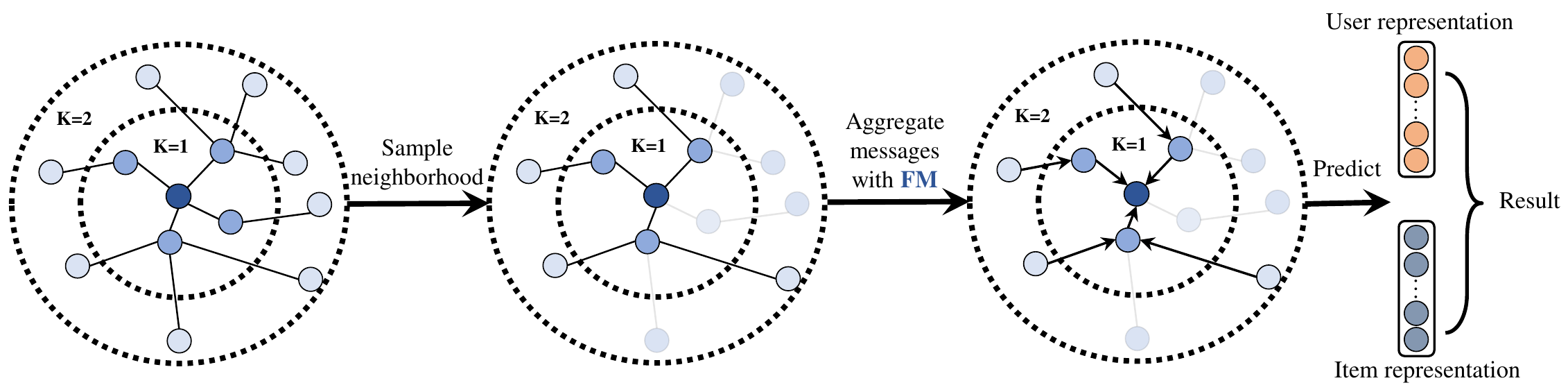}
\caption{The proposed \modela~model.}
\label{fig:modela}%
\end{figure*}%

In this section, we first formulate the problem, then we present the details of the proposed model \modela~and the cross-domain framework as shown in Figure \ref{fig:modela} and \ref{fig:modelb}, respectively.

\subsection{Problem Formulation}
Assuming $\setU=\{u_1, u_2,... ,u_{|U|}\}$ is a set of $|U|$ users, $\setI=\{i_1,i_2... ,i_{|I|}\}$ is a set of $|I|$ items. There is now a source domain $\D_s$ containing a user set $\setU_s$ and an item set $\setI_s$, and a target domain $\D_t$ containing a user set $\setU_t$ and an item set $\setI_t$. In each domain, there is a explicit or implicit feedback matrix from users and items (such as rating, watching, clicking, buying, etc.). 
The task of single-domain recommendation is to improve the recommendation performance on the target domain $\D_t$ via utilizing its feedback matrix.
Now assuming some information is shared between these two domains, such as overlapped users or items. The task of cross-domain recommendation is to combine the data and knowledge of the source domain $\D_s$ to help improve the recommendation performance of the target domain $\D_t$.

\subsection{Graph Factorization Machines}\label{sec:GFM}
In this subsection, we present the proposed \modela~as illustrated in Figure \ref{fig:modela}. The \modela~first samples neighbors for each node, then aggregates messages from neighbors via the Equation (\ref{equ:nfm}) and finally makes prediction with the aggregated messages.

\subsubsection{Factorization Machines} 
In recommender systems, the users and items features usually are one-hot encoding categorical features, the dimension is usually high and the vectors are sparse.
FM is an effective method to address such high-dimension and sparse problems and it can be seen as performing the latent relation modeling between any two features. In the graph-structure scenario we focus on, the features can be node features, e.g., user or item nodes.

Firstly, we project each non-zero node $n$ to a low dimension dense vector representation $\v_n$.
Embedding is a popular solution in the neural network over various application scenarios.
It learns one embedding vector $\v_n\in\setR^{k}$ for each node $n$, where $k$ is the dimension of embedding vectors. 

Different from the traditional FM, which uses inner product to get a scalar, in neural network, we need to get a vector representation via the Hadamard product as done in \cite{he2017nfm}:
\begin{eqnarray}
FM(\x) = \sum_{n\neq m} x_n\v_n\odot x_m\v_m,\label{equ:fm}
\end{eqnarray}
where $x_n, x_m\in\{0, 1\}$ indicate the presence or absence of the node $n$ and $m$ as shown in Figure \ref{fig:example},
The Hadamard product $\odot$ denotes the element-wise product of two vectors:
\begin{eqnarray}
(\v_n\odot\v_m)_k=v_{nk}v_{mk}.
\end{eqnarray}
The computing complexity of the above Equation (\ref{equ:fm}) is in $O(k|V|^2)$, where $|V|$ is the number of nodes in $\x$, since all pairwise relations need to be computed.
Actually, the Hadamard product based FM can be reformulated to linear runtime $O(k|V|)$ \cite{he2017nfm} just like the inner product based FM \cite{rendle2010factorization} as follows:
\begin{eqnarray}
FM(\x) = \frac{1}{2}\left[(\sum_{n} x_n\v_n)^2- \sum_{n} (x_n\v_n)^2\right],
\end{eqnarray}
where $\v^2$ denotes $\v\odot\v$.
Besides, in sparse settings, the sums only need to be computed over the non-zero pairs $x_n x_m$. Therefore, the actual computing complexity is $O(k\hat{|V|})$, where $\hat{|V|}$ is the number of non-zero nodes in $\x$.
By adding the MLP, the FM can model higher-order feature interactions by \cite{he2017nfm}:
\begin{eqnarray}
\hat{FM}(\x) = MLP(FM(\x)),\label{equ:nfm}
\end{eqnarray}

\subsubsection{Graph Factorization Machine Layer}
While FM is an effective method to address high-dimension and sparse features in recommender systems, it is not designed for graph-structure data and cannot consider the graph topology information, i.e., the multi-order neighbor information. Here, we extend the FM to the graph-structure data to form a new GNN model, which is more suitable for recommender systems. 
Similar with the GraphSAGE \cite{hamilton2017graphsage}, the \modela~ parameters can also be learned using standard stochastic gradient descent and backpropagation techniques.

\textbf{Sampling Neighborhood.} In this work, we first uniformly sample a fixed-size set $\N(n)$ of neighbors for each node $n$.
If the neighbor number of node $n$ is large than the sampling threshold $\delta$, the sampling without replacement is used, otherwise the sampling with replacement is used.
It should be mentioned that designing a different neighbor sampling scheme is not the focus of this paper as we aim at designing a powerful message aggregator to learn efficient node representations.
In fact, any advanced neighbor sampling scheme can be easily
integrated into our framework, making the proposed \modela~general and flexible.

\textbf{Aggregating Messages.}
Most of the prevailing GNN models follow the neighborhood aggregation strategy and are analogous to Weisfeiler-Lehman
(WL) graph isomorphism test \cite{xu2018powerful}.
The representation of a node is obatined by iteratively aggregating messages from its neighbors.
We adopt the Equation (\ref{equ:nfm}) as the neighborhood aggregator in graph neural networks, the node representation $\h^l_n$ in the $l$-th layer of node $n$ is relevant to itself and its neighbor representations in the $(l-1)$-th layer.
Note that the node $n$ can be a user or an item.
\begin{eqnarray}
\h^l_n&=&AGGR(\{\h^{l-1}_n,\h^{l-1}_m  \big| \forall m\in\N(n)\})\nonumber\\
&=&\hat{FM}(\N(n)).\label{equ:aggregator}
\end{eqnarray}
Note that we add self-loop for all nodes before sampling neighborhood, so the node itself may be sampled in the $\N(n)$.
By stacking multiple \modela~layers, the message aggregator can capture higher-order neighbor messages.
Specifically, for the user-item feedback graph $G$ (i.e., if a user has feedback on an item, there is an edge between the user node and the item node), the \modela~uses the same aggregation scheme for all users and items.
For example, if the user $u$ has feedback on sampled items $\{i_1,...,i_\delta\}$, the aggregator can model the feature interactions among $\{i_1,...,i_\delta\}$ for obtaining the representation of user $u$.
Similarly, if sampled users $\{u_1,...,u_\delta\}$ have feedback on the item $i$, the aggregator can model the feature interactions among $\{u_1,...,u_\delta\}$ for obtaining the representation of item $i$.
By stacking \modela~layers, the representations of users and items can be iteratively aggregated from their multi-order (multi-hop) neighbors.

\textbf{Making prediction.}
To predict the interaction probability between a given pair of user and item$(u,i)$,
we adopt a simple but widely-used inner product predictive function to estimate $\hat{\y}_{ui}$.
The inner product acts on the user representation $\h_u$ and item representation $\h_i$ learned via the \modela~according to the Equation (\ref{equ:aggregator}):
\begin{eqnarray}
\hat{\y}_{ui}=\sigma(\h_u^{\top}\h_i),
\end{eqnarray}
where $\sigma$ is the \textit{sigmoid} function.

In our implicit feedback recommendation scenario, we can observe the implicit interactions between users and items. Thus, to train an end-to-end \modela, we use the negative logarithm of joint probability as the loss function (i.e., \textit{logloss}), which is widely used to optimize implicit feedback recommendation tasks \cite{mnih2008probabilistic,kabbur2013fism,he2017neural,gao2019natr,hu2019hybrid}:
\begin{eqnarray}
\L(\theta)=-(\sum_{(u,i)\in\Y_{+}}\log\hat{\y}_{ui}+\sum_{(u,i)\in\Y_{-}}\log(1-\hat{\y}_{ui})),\label{loss}
\end{eqnarray}
where $\Y_{+}$ denotes the set of observed implicit feedback, and $\Y_{-}$ denotes the set of negative samples sampled from unobserved implicit feedback. $\theta$ is the parameter set which contains the all embedding vectors $\v_n$ and the parameters in the MLP in Equation (\ref{equ:nfm}).

To construct a mini-batch, we follow the existing works \cite{gao2019natr,hu2019hybrid} to first sample a batch of user-item interaction paris $(u,i)$. For each $(u,i)$, we then adopt negative sampling to randomly select unobserved items $\{i_{1}^{'},...,i_{\gamma}^{'}\}$ for user $u$ with a sampling ratio of $\gamma$. Finally, we obtain $\gamma$ triplets $\{(u,i,i_{1}^{'}),...,(u,i,i_{\gamma}^{'})\}$ for each user in a batch.
Note that we do not perform a predefined negative sampling in advance since this can only generate a fixed training set of negative samples. Instead, we generate negative samples during each epoch, enabling diverse and augmented training sets of negative examples to be used \cite{hu2019hybrid}.

\subsection{General Cross-Domain Framework}
\begin{figure*}[!t]
\begin{center}
\includegraphics[width=0.85\linewidth]{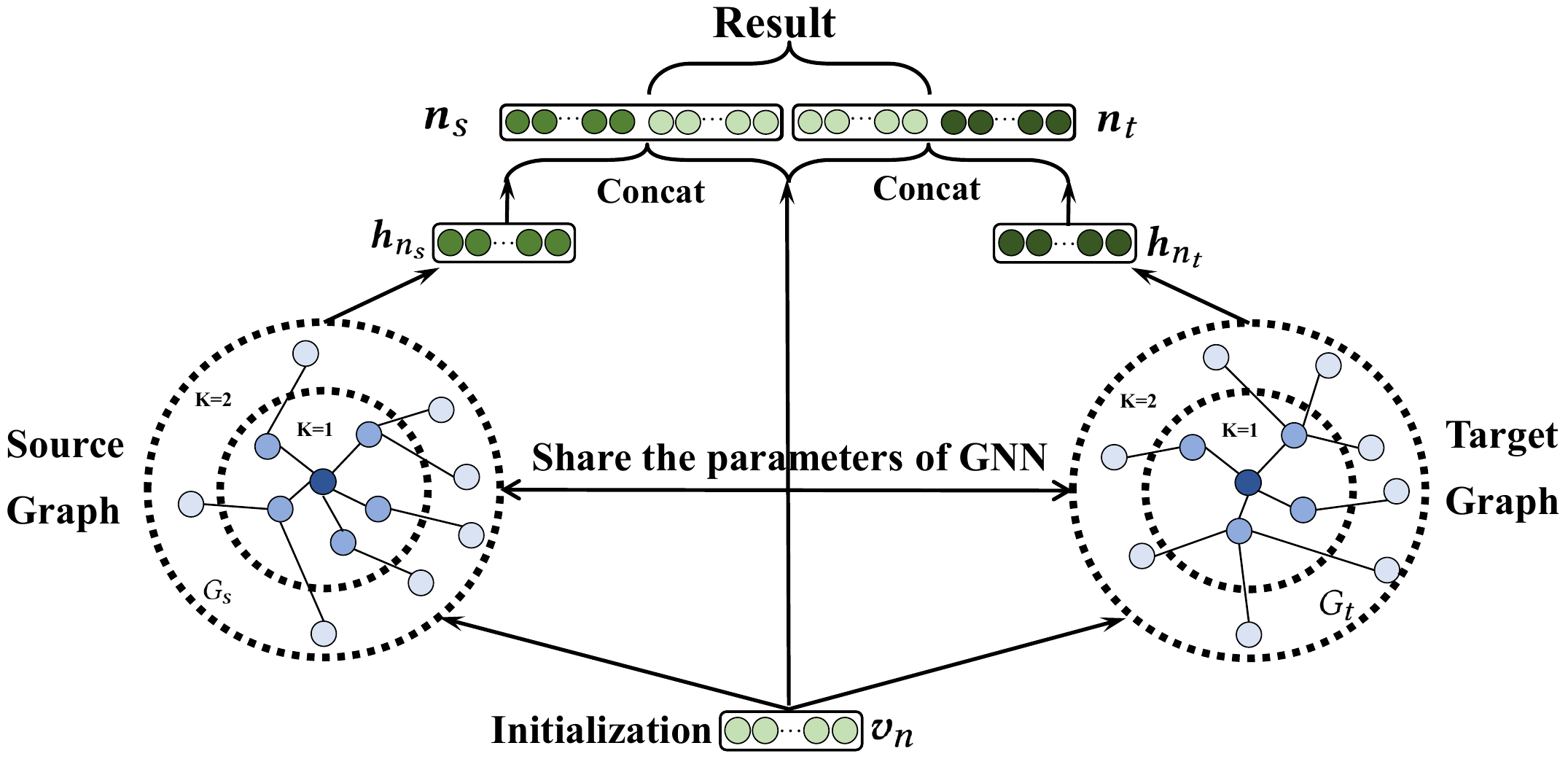}
\caption{The proposed general cross-domain framework. For shared nodes, we initialize the node representations as $\v_n$, for unshared nodes, the $\v_n$ represents $\v_{n_s}$ and $\v_{n_t}$ in the source and target domains, respectively.}
\label{fig:modelb}%
\end{center}
\end{figure*}%
In this subsection, we present the proposed cross-domain recommendation framework as illustrated in Figure \ref{fig:modelb} and apply the framework to other GNN models.

\subsubsection{Cross-Domain Graph Factorization Machines}
As mentioned before, cross-domain recommendation \cite{pan2010cst,tang2012ctl,man2017emcdr,chen2019eatnn,hu2019hybrid,yuan2019darec} is an effective method to leverage auxiliary data from other domains to improve the data sparsity and recommendation quality of the target domain.
In order to apply the proposed \modela~to cross-domain recommendation on the graph-structured data, we propose a general cross-domain recommendation framework which can be applied not only to the proposed \modela~to form the cross-domain GFM (\modelb), but also to other GNN models.

First, for the graphs $G_s$ and $G_t$ in the source domain $\D_s$ and target domain $\D_t$, respectively, we assume that $G_s$ and $G_t$ have some shared nodes, which can be users or items, or both in the user-item feedback graphs.
For these shared nodes, we initialize the node representations $\v_n$ in Equation (\ref{equ:fm}) with the same embedding vectors in both source and target domains. These representations can be seen as the domain-shared features and learned automatically during the training stage by collaborating  with the \modela~in their respective domains.
For unshared nodes, we initialize the node representations $\v_n$ in Equation (\ref{equ:fm}) with different embedding vectors $\v_{n_s}$ and $\v_{n_t}$ in source and target domains.

Then, the source and target domains use the graph-structure data in their respective domains for multi-layer \modela~learning.
This is equivalent to the fact that nodes in each domain learn the domain-specific representations based on the topology information in their own domains.
Besides, in order to further integrate the knowledge in the two domains, the \modela~in the two domains can learn the node representation cooperatively via sharing the parameters in the MLP in Equation (\ref{equ:nfm}).
Thus, we obtain the domain-specific node representations $\h_{n_s}$ and $\h_{n_t}$ of both source and target domains simultaneously.

Next, the domain-shared and domain-specific representations in each domain are combined together as the final node representations:
\begin{eqnarray}
\n_s&=&[\v_{n_s};\h_{n_s}],\label{concats}\label{concat1}\\
\n_t&=&[\v_{n_t};\h_{n_t}],\label{concatt}\label{concat2}
\end{eqnarray}
where $\v_{n_s}$ and $\v_{n_t}$ are same when the node $n$ is shared in the source and target domains, and $[\cdot;\cdot]$ denotes the concatenation of two node representations.

Finally, for cross-domain recommendation tasks, the framework can be design in an end-to-end scheme to make prediction via inner product between a given pair of user and item $(u,i)$ in each domain:
\begin{eqnarray}
\hat{\y}_{u_si_s}&=&\sigma(\n_{u_s}^{\top}\n_{i_s}),\\
\hat{\y}_{u_ti_t}&=&\sigma(\n_{u_t}^{\top}\n_{i_t}),
\end{eqnarray}
where $\n_{u_s}$ and $\n_{u_t}$ are the user node representations in the source and target domains, respectively, and $\n_{i_s}$ and $\n_{i_t}$ are the item node representations in the source and target domains, respectively.
They are all learned from Equations (\ref{concats}) and (\ref{concatt}).
The $\sigma$ is the \textit{sigmoid} function.

The loss function $\L_{st}(\theta)$ of the \modelb~combines two components to a unified multi-task learning framework:
\begin{eqnarray}
\L_{st}(\theta)=\alpha\L_s(\theta_s)+(1-\alpha)\L_t(\theta_t)\label{equ:lossst}
\end{eqnarray}
where $\theta_s$ and $\theta_t$ are the parameter sets in the source and target domains, respectively. The $\L_s(\cdot)$ and $\L_t(\cdot)$ are the loss functions defined in the Equation (\ref{loss}). The tunable hyper-parameter $\alpha$ controls the different strength of the two components.

The mini-batch construction and training scheme are the same as the single \modela~model as described in the Section \ref{sec:GFM}.

\subsubsection{Apply the Framework to Other GNN Models}\label{sec:general}
The general cross-domain framework as shown in Figure \ref{fig:modelb} could be applied upon various existing GNN models. The most important thing is to apply the framework to define the domain-shared and domain-specific representations.
For our \modelb, the domain-shared representations are learned from the initialized shared node representations, which is consistent with other GNN models. So in other GNN models, the shared nodes should be randomly initialized as the same vectors. 
The domain-specific representations are learned from the graph topology information in each domain based on the respective GNN models.
Besides, for our \modelb, the model learns the node representation cooperatively via sharing the parameters in the MLP in Equation (\ref{equ:nfm}).
For other GNN models, the parameters within the each GNN model can be shared for further integrating the knowledge in the two domains.
The loss function and training process are consistent with the \modelb.
Based on the above strategies, the proposed general cross-domain framework can be applied to GCN \cite{kipf2016gcn}, GAT \cite{velivckovic2017gat}, GraphSAGE \cite{hamilton2017graphsage} and so on.

\section{Experiments}
\begin{table*}[!t]
  \centering
  \caption{Statistics of the datasets. ``$\#$'' means the number of the corresponding items.}
  \resizebox{.75\linewidth}{!}{
    \begin{tabular}{llllll}
    \toprule
    \multirow{2}[4]{*}{Dataset} & \multirow{2}[4]{*}{Shared (\#)} & \multicolumn{2}{c}{Source Domain} & \multicolumn{2}{c}{Target Domain} \\
\cmidrule{3-6}          &       & Unshared (\#) & \#Feedback & Unshared (\#) & \#Feedback \\
    \midrule
    TC$\rightarrow$IQI & Item (5,568) & User (35,398) & 314,621 & User (19,999) & 78,429 \\
    ML$\rightarrow$NF & Item (5,565) & user (30,279) & 11,555,621 & User (11,498) & 199,765 \\
    MO$\rightarrow$MU & User (27,898) & Item (15,465) & 7,366,992 & Item (14,521) & 3,784,331 \\
    MU$\rightarrow$BO & User (27,898) & Item (14,521) & 3,784,331 & Item (15,774) & 1,936,754 \\
    \bottomrule
    \end{tabular}%
    }
  \label{tab:dataset}%
\end{table*}%
In this section, we perform experiments to evaluate the proposed model and framework against various baselines on real-world datasets. 
We first introduce the datasets, evaluation protocol, implementation details and baseline methods of our experiments. Finally, we present our experimental results and analysis.

\subsection{Datasets}
We utilize four pairs frequently used real-world datasets, which contain two pairs \textbf{user-shared} datasets and two pairs \textbf{item-shared} datasets. 
For all datasets, we only use the user IDs, item IDs and their implicit feedback information.
For simplicity, we intentionally transform the rating data into binary (1/0, indicating whether a user has interacted with an item or not) to fit the problem setting of implicit feedback following \cite{gao2019natr}.
The statistics of the four pairs datasets are listed in Table \ref{tab:dataset}.
\begin{itemize}
\item \textbf{TC$\rightarrow$IQI} \cite{yan2019tciqi} are from two mainstream video websites Tencent (TC)\footnote{https://v.qq.com} and iQIYI (IQI)\footnote{https://www.iqiyi.com} in China. 
There are a lot of overlapped items (movies) in the two websites.  
We take TC and IQI as the source and target domains, respectively. 
We got the processed dataset pair directly from \cite{yan2019tciqi}.

\item \textbf{ML$\rightarrow$NF}\footnote{https://grouplens.org/datasets/movielens}$^,$\footnote{https://www.kaggle.com/laowingkin/netflix-movie-recommendation/data} are from two popular movie recommendation platforms MovieLens and Netflix, in which there are a lot of overlapped items (movies). 
We take MovieLens (ML) as the source domain and the Netflix (NF) as the target domain. We identify the same movies with their names (case insensitive) and years to avoid wrong identifications as possible, which is similar data processing method with \cite{gao2019natr}.

\item \textbf{MO$\rightarrow$MU} are from the famous social network platform Douban\footnote{https://www.douban.com\label{douban}} in China. Overlapped users have feedback on both Movie (MO) and Music (MU).
We take MO as the source domain and the MU as the target domain.

\item \textbf{MU$\rightarrow$BO} are also from the famous social network platform Douban\textsuperscript{\ref{douban}} in China. Overlapped users have feedback on both Music (MU) and Book (BO).
We take MU as the source domain and the BO as the target domain.
\end{itemize}
\begin{table*}[!t]
  \centering
  \caption{The experimental results evaluated by HR@K and NDCG@K on single domain recommendation task with 95\% confidence intervals.}
  \resizebox{1\linewidth}{!}{
    \begin{tabular}{clcccccc}
    \toprule
    Dataset & Model & HR(NDCG)@1 & HR@10 & HR@50 & NDCG@10 & NDCG@50 & \textbf{Average} \\
    \midrule
    \multirow{6}[3]{*}{IQI} & NCF   & 0.1545$\pm$0.0029 & 0.5004$\pm$0.0039 & 0.9153$\pm$0.0015 & 0.2986$\pm$0.0020 & 0.4185$\pm$0.0088 & 0.4575  \\
          & GCN   & 0.0877$\pm$0.0040 & 0.4747$\pm$0.0233 & 0.6620$\pm$0.0323 & 0.2937$\pm$0.0116 & 0.3361$\pm$0.0137 & 0.3708  \\
          & GAT   & 0.1497$\pm$0.0545 & \textbf{0.5878$\pm$0.0765} & 0.9589$\pm$0.0100 & 0.3359$\pm$0.0797 & 0.4368$\pm$0.0632 & 0.4938  \\
          & GraphSAGE-mean & 0.0912$\pm$0.0243 & 0.5671$\pm$0.0388 & 0.9618$\pm$0.0013 & 0.3145$\pm$0.0298 & 0.3943$\pm$0.0234 & 0.4658  \\
          & GraphSAGE-pooling & 0.1122$\pm$0.0217 & 0.5796$\pm$0.0522 & 0.9508$\pm$0.0041 & 0.3083$\pm$0.0346 & 0.3956$\pm$0.0231 & 0.4693  \\
\cmidrule{2-8}          & \textbf{GFM} & \textbf{0.1591$\pm$0.0278} & 0.5821$\pm$0.0486 & \textbf{0.9671$\pm$0.0060} & \textbf{0.3376$\pm$0.0315} & \textbf{0.4391$\pm$0.0228} & \textbf{0.4970 } \\
\midrule
    \multirow{6}[3]{*}{NF} & NCF   & 0.2102$\pm$0.0038 & 0.5840$\pm$0.004 & 0.8706$\pm$0.0025 & 0.3804$\pm$0.0036 & 0.4446$\pm$0.0034 & 0.4980  \\
          & GCN   & 0.1048$\pm$0.0141 & 0.1688$\pm$0.0141 & 0.4981$\pm$0.0212 & 0.1328$\pm$0.0144 & 0.2009$\pm$0.0159 & 0.2211  \\
          & GAT   & 0.1918$\pm$0.0045 & 0.5564$\pm$0.0027 & 0.9028$\pm$0.0030 & 0.3554$\pm$0.0021 & 0.4318$\pm$0.0026 & 0.4876  \\
          & GraphSAGE-mean & 0.1920$\pm$0.0053 & 0.5525$\pm$0.0008 & 0.8874$\pm$0.0025 & 0.3542$\pm$0.0025 & 0.4280$\pm$0.0030 & 0.4828  \\
          & GraphSAGE-pooling & 0.2059$\pm$0.0027 & 0.6054$\pm$0.0034 & \textbf{0.9217$\pm$0.0014} & 0.3906$\pm$0.0027 & \textbf{0.4696$\pm$0.0023} & \textbf{0.5186 } \\
\cmidrule{2-8}          & \textbf{GFM} & \textbf{0.2140$\pm$0.0042} & \textbf{0.6077$\pm$0.0131} & 0.9184$\pm$0.0054 & \textbf{0.3918$\pm$0.0072} & 0.4613$\pm$0.0055 & \textbf{0.5186 } \\
    \midrule
    \multirow{6}[3]{*}{MU} & NCF   & 0.2046$\pm$0.0043 & 0.6078$\pm$0.0026 & \textbf{0.9590$\pm$0.0007} & 0.3835$\pm$0.0036 & \textbf{0.5093$\pm$0.0031} & 0.5328  \\
          & GCN   & 0.1594$\pm$0.0002 & 0.4984$\pm$0.0019 & 0.7589$\pm$0.0034 & 0.2946$\pm$0.0006 & 0.3981$\pm$0.0008 & 0.4219  \\
          & GAT   & 0.2335$\pm$0.0159 & 0.6833$\pm$0.0072 & 0.9545$\pm$0.0005 & 0.4463$\pm$0.0128 & 0.5002$\pm$0.0112 & 0.5636  \\
          & GraphSAGE-mean & 0.1927$\pm$0.0121 & 0.5923$\pm$0.0196 & 0.8901$\pm$0.0220 & 0.3742$\pm$0.0161 & 0.4406$\pm$0.0167 & 0.4980  \\
          & GraphSAGE-pooling & 0.2215$\pm$0.0193 & 0.6210$\pm$0.0190 & 0.9484$\pm$0.0026 & 0.4145$\pm$0.0208 & 0.4965$\pm$0.0171 & 0.5404  \\
\cmidrule{2-8}          & \textbf{GFM} & \textbf{0.2399$\pm$0.0026} & \textbf{0.6887$\pm$0.0009} & 0.9507$\pm$0.0028 & \textbf{0.4470$\pm$0.0011} & 0.5055$\pm$0.0028 & \textbf{0.5664 } \\
    \midrule
    \multirow{6}[3]{*}{BO} & NCF   & 0.2567$\pm$0.0081 & 0.6733$\pm$0.007 & 0.9422$\pm$0.0024 & 0.4558$\pm$0.0081 & 0.5164$\pm$0.007 & 0.5689  \\
          & GCN   & 0.1899$\pm$0.0004 & 0.5007$\pm$0.0017 & 0.6991$\pm$0.001 & 0.3558$\pm$0.0002 & 0.3900$\pm$0.0002 & 0.4271  \\
          & GAT   & 0.2805$\pm$0.0258 & 0.7034$\pm$0.0365 & 0.9369$\pm$0.0202 & \textbf{0.4776$\pm$0.0321} & 0.5303$\pm$0.0286 & 0.5857  \\
          & GraphSAGE-mean & 0.2137$\pm$0.0009 & 0.6036$\pm$0.0007 & 0.8741$\pm$0.0022 & 0.3920$\pm$0.0007 & 0.4525$\pm$0.001 & 0.5072  \\
          & GraphSAGE-pooling & 0.2716$\pm$0.0148 & 0.6987$\pm$0.0143 & 0.9351$\pm$0.0051 & 0.4653$\pm$0.0155 & 0.5166$\pm$0.0136 & 0.5775  \\
\cmidrule{2-8}          & \textbf{GFM} & \textbf{0.2867$\pm$0.005} & \textbf{0.7055$\pm$0.0063} & \textbf{0.9431$\pm$0.0042} & 0.4757$\pm$0.0061 & \textbf{0.5392$\pm$0.0058} & \textbf{0.5900 } \\
    \bottomrule
    \end{tabular}%
    }
  \label{tab:single}%
\end{table*}%

\subsection{Evaluation Protocol}
Following existing works \cite{he2017neural,hu2019hybrid}, we adopt the Leave-One-Out (LOO) evaluation.
We randomly sample one interaction for each user as the validation and test sets, respectively.
We also follow the common strategy \cite{hu2019hybrid,gao2019natr} to randomly sample 99 unobserved (negative) items for each user and then evaluate how well the model can rank the test item against these negative ones. 
Then, we adopt two standard metrics, \textbf{HR@K} and \textbf{NDCG@K}, which are widely used in recommendation \cite{gao2019natr,hu2019hybrid,he2017neural,wang2018tem,ding2018improving}, to evaluate the ranking performance of each methods. The HR@K is computed as follows:
\begin{eqnarray}
HR@K=\frac{1}{|U|}\sum_{u\in\setU} I(p_u\leq K),
\end{eqnarray}
where $p_u$ is the hit position for the user $u$'s test item, and $I(\cdot)$ is the indicator function.
The NDCG@K is computed as follows:
\begin{eqnarray}
NDCG@K=\frac{1}{|U|}\sum_{u\in\setU} \frac{\log 2}{\log (p_u+1)}.
\end{eqnarray}
We report HR@K and NDCG@K with K = 1, 10 and 50.
The larger the value, the better the performance for all the evaluation metrics.
For all experiments, we report the metrics with \textbf{95\%} \textbf{confidence intervals} on five runs.

\subsection{Implementation Details}
If a user has feedback on an item, there is an edge between the user node and the item node.
Thus, we construct the feedback graph $G$ utilized in our experiments.

For single domain recommendation task, we perform experiments on the four target domain datasets (i.e., IQI, NF, MU, BO).
For all datasets we use: embedding dimension $k=32$, neighbor sampling threshold $\delta=30$ with two \modela~layers, negative sampling ratio $\gamma=8$, mini-batch size of 256 and learning rate of 0.001. 
We also use dropout whose probability is 0.4.

For cross-domain recommendation task, we perform experiments on the four pairs cross-domain datasets.
For all datasets we use: embedding dimension $k=16$, neighbor sampling threshold $\delta=10$ with one \modela~layer, negative sampling ratio $\gamma=8$,
tunable hyper-parameter $\alpha=0.7$ to control the different strength in Equation (\ref{equ:lossst}),
mini-batch size of 256 and learning rate of 0.001. 
We also use dropout whose probability is 0.4.

All these values and hyper-parameters of all baselines are chosen via a grid search on the IQI validation set.
We do not perform any datasets-specific tuning except early stopping on validation sets.
All models are implemented using TensorFlow\footnote{https://www.tensorflow.org} and trained on GTX 1080ti GPU.
Training is finished through stochastic gradient descent over shuffled mini-batches with the Adam \cite{kingma2014adam} update rule.

\subsection{Baseline Methods}
We construct three groups of experiments to demonstrate the effectiveness of the proposed model and framework.
\subsubsection{Single Domain Recommendation}
We compare the proposed \modela~model with the following baseline models.
\begin{itemize}
\item \textbf{NCF}~\cite{he2017neural}: Neural Collaborative Filtering (NCF) is the state-of-the-art solution for recommendation tasks with implicit feedback.
We use one of the variants of NCF, which is also called Generalized Matrix Factorization (GMF).
\item \textbf{GCN}~\cite{kipf2016gcn}: The vanilla GCN learns latent node representations based on the first-order approximation of spectral graph convolutions. 
\item \textbf{GAT}~\cite{velivckovic2017gat}: It applies the attention mechanism to learn different weights for aggregating node features from neighbors. 
\item \textbf{GraphSAGE-mean}~\cite{hamilton2017graphsage}: It learns to aggregate node messages from a node’s local neighborhood by the mean aggregator.
\item \textbf{GraphSAGE-pooling}~\cite{hamilton2017graphsage}: It learns to aggregate node messages from a node’s local neighborhood by the pooling aggregator.
\end{itemize}
For GCN, GAT, GraphSAGE-mean and GraphSAGE-pooling, We apply the inner product on the user and item node representations as the output.

\subsubsection{Cross-Domain Recommendation}
We compare the proposed \modelb~model with the following baseline models.
\begin{itemize}
\item \textbf{CST}~\cite{pan2010cst}: Coordinate System Transfer (CST) assumes that both users and items are overlapped and adds
two regularization terms in its objective function.  Here, we adapt the CST to our datasets by only reserving single-side (i.e., the user-side or item-side) regularization term.
\item \textbf{CD-NCF}~\cite{he2017neural}: Neural Collaborative Filtering (NCF) is the state-of-the-art solution for single domain recommendation tasks with implicit feedback. Here, we adapt it to our cross-domain recommendation task via sharing the overlapped user or item embeddings.
\item \textbf{EMCDR}~\cite{man2017emcdr}: This is an embedding and mapping framework for cross-domain recommendation.
The framework contains Latent Factor Model, Latent Space Mapping and Cross-domain Recommendation, and it is not an end-to-end method.
\item \textbf{EATNN}~\cite{chen2019eatnn}: This is the state-of-the-art solution for cross-domain recommendation tasks. By introducing attention mechanisms, the model automatically assigns a personalized transfer scheme for each user.
\end{itemize}

\subsubsection{General Cross-Domain Recommendation}
We apply the proposed cross-domain framework to other baseline GNN models.
\begin{itemize}
\item \textbf{CD-GCN}~\cite{kipf2016gcn}: It applies the proposed general framework to the GCN as described in Section \ref{sec:general}. 
\item \textbf{CD-GAT}~\cite{velivckovic2017gat}: It applies the proposed general framework to the GAT. 
\item \textbf{CD-GraphSAGE-mean}~\cite{hamilton2017graphsage}: It applies the proposed general framework to the GraphSAGE-mean. 
\item \textbf{CD-GraphSAGE-pooling}~\cite{hamilton2017graphsage}: It applies the proposed general framework to the GraphSAGE-pooling. 
\end{itemize}

\subsection{Performance Comparison}
\begin{table*}[!t]
  \centering
  \caption{The experimental results evaluated by HR@K and NDCG@K on cross-domain recommendation task with 95\% confidence intervals.}
  \resizebox{1\linewidth}{!}{
    \begin{tabular}{clcccccc}
    \toprule
    Dataset & Model & HR(NDCG)@1 & HR@10 & HR@50 & NDCG@10 & NGDCG@50 & \textbf{Average} \\
    \midrule
    \multirow{5}[3]{*}{TC$\rightarrow$IQI} & CST   & 0.1948$\pm$0.0039 & \textbf{0.6678$\pm$0.0136} & 0.9455$\pm$0.0028 & 0.4178$\pm$0.0099 & 0.4858$\pm$0.0030 & 0.5423  \\
          & CD-NCF & 0.1701$\pm$0.0314 & 0.5408$\pm$0.0445 & 0.8702$\pm$0.0402 & 0.3392$\pm$0.0411 & 0.4131$\pm$0.0396 & 0.4667  \\
          & EMCDR & 0.2058$\pm$0.0239 & 0.3962$\pm$0.0628 & 0.7438$\pm$0.0436 & 0.2897$\pm$0.0394 & 0.3640$\pm$0.0358 & 0.3999  \\
          & EATNN & 0.1959$\pm$0.0102 & 0.6473$\pm$0.0089 & 0.9314$\pm$0.0026 & 0.4103$\pm$0.0100 & 0.4906$\pm$0.0087 & 0.5351  \\
\cmidrule{2-8}          & \textbf{CD-GFM} & \textbf{0.2105$\pm$0.0089} & 0.6536$\pm$0.0159 & \textbf{0.9758$\pm$0.0088} & \textbf{0.4222$\pm$0.0108} & \textbf{0.4963$\pm$0.0080} & \textbf{0.5517 } \\
\midrule
    \multirow{5}[3]{*}{ML$\rightarrow$NF} & CST   & 0.1878$\pm$0.0058 & 0.5413$\pm$0.0024 & 0.8551$\pm$0.0007 & 0.3486$\pm$0.0015 & 0.4178$\pm$0.0023 & 0.4701  \\
          & CD-NCF & 0.1997$\pm$0.0260 & 0.5540$\pm$0.0457 & 0.8539$\pm$0.0246 & 0.3600$\pm$0.0353 & 0.4266$\pm$0.0310 & 0.4788  \\
          & EMCDR & 0.0968$\pm$0.0260 & 0.3406$\pm$0.0240 & 0.6522$\pm$0.0730 & 0.2027$\pm$0.0170 & 0.2708$\pm$0.0070 & 0.3126  \\
          & EATNN & 0.2103$\pm$0.0018 & 0.5892$\pm$0.0038 & 0.8745$\pm$0.0016 & 0.3835$\pm$0.0015 & 0.4472$\pm$0.0013 & 0.5009  \\
\cmidrule{2-8}          & \textbf{CD-GFM} & \textbf{0.2243$\pm$0.0047} & \textbf{0.6247$\pm$0.0069} & \textbf{0.9228$\pm$0.0033} & \textbf{0.4062$\pm$0.0055} & \textbf{0.4732$\pm$0.0043} & \textbf{0.5302 } \\
    \midrule
    \multirow{5}[3]{*}{MO$\rightarrow$MU} & CST   & 0.2378$\pm$0.0085 & 0.5934$\pm$0.0024 & 0.9051$\pm$0.0073 & 0.3986$\pm$0.0115 & 0.4775$\pm$0.0035 & 0.5225  \\
          & CD-NCF & 0.2599$\pm$0.0200 & 0.7232$\pm$0.0430 & 0.9480$\pm$0.0261 & 0.4747$\pm$0.0315 & 0.5281$\pm$0.0281 & 0.5868  \\
          & EMCDR & 0.2290$\pm$0.0290 & 0.5610$\pm$0.0703 & 0.8430$\pm$0.0560 & 0.3834$\pm$0.0320 & 0.4234$\pm$0.0410 & 0.4880  \\
          & EATNN & 0.2680$\pm$0.0021 & 0.7253$\pm$0.0035 & 0.9457$\pm$0.0026 & \textbf{0.4881$\pm$0.0013} & 0.5282$\pm$0.0014 & 0.5911  \\
\cmidrule{2-8}          & \textbf{CD-GFM} & \textbf{0.2728$\pm$0.0054} & \textbf{0.7314$\pm$0.0072} & \textbf{0.9671$\pm$0.002} & 0.4851$\pm$0.0060 & \textbf{0.5389$\pm$0.0049} & \textbf{0.5991 } \\
    \midrule
    \multirow{5}[3]{*}{MU$\rightarrow$BO} & CST   & 0.2524$\pm$0.0089 & 0.6973$\pm$0.0102 & 0.9355$\pm$0.0098 & 0.4575$\pm$0.0105 & 0.5143$\pm$0.0068 & 0.5714  \\
          & CD-NCF & 0.2770$\pm$0.0158 & 0.7184$\pm$0.0332 & \textbf{0.9472$\pm$0.0261} & 0.4841$\pm$0.0215 & 0.5334$\pm$0.0836 & 0.5920  \\
          & EMCDR & 0.2004$\pm$0.2972 & 0.4864$\pm$0.5881 & 0.7612$\pm$0.4115 & 0.3324$\pm$0.4423 & 0.3920$\pm$0.4082 & 0.4345  \\
          & EATNN & 0.2731$\pm$0.0015 & 0.7064$\pm$0.0036 & 0.9277$\pm$0.0026 & 0.4634$\pm$0.0013 & 0.5070$\pm$0.0017 & 0.5755  \\
\cmidrule{2-8}          & \textbf{CD-GFM} & \textbf{0.2978$\pm$0.0481} & \textbf{0.7267$\pm$0.0688} & 0.9424$\pm$0.0295 & \textbf{0.4872$\pm$0.0609} & \textbf{0.5502$\pm$0.0523} & \textbf{0.6009 } \\
    \bottomrule
    \end{tabular}%
    }
  \label{tab:cross}%
\end{table*}%
\begin{figure*}[!t]
\begin{center}
\includegraphics[width=\linewidth]{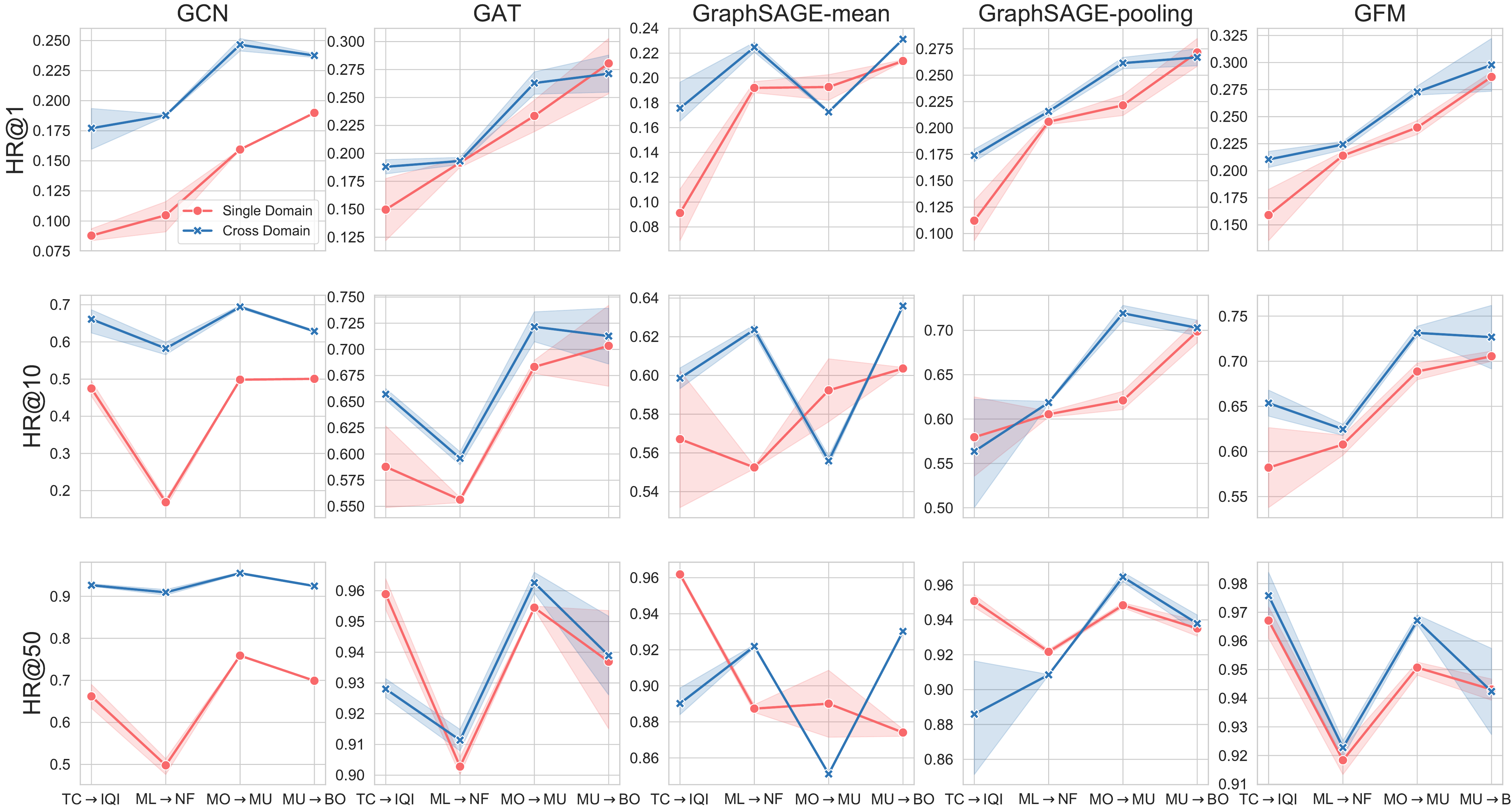}
\caption{The HR@K results of the general cross-domain framework on 4 (datastes) $\times$ 10 (models) = 40 tasks.}
\label{general_framework}%
\end{center}
\end{figure*}%
\subsubsection{Single Domain Recommendation Task}
We demonstrate the effectiveness of our \modela~on four target domain datasets.
The experimental results evaluated by HR@K and NDCG@K on IQI, NF, MU and Bo are presented in Table \ref{tab:single}. 
From these results, we have the following insightful observations.
\begin{itemize}
	\item[-] Among these GNN baselines, the GCN has acceptable performances on multiple datasets. 
	The GraphSAGE-mean improves the results comparing with GCN via introducing the mean aggregator to aggregate messages from each node's local neighborhood. 
	The GraphSAGE-pooling achieves further improvement over GraphSAGE-mean by replacing the mean aggregator with the more complex pooling aggregator, which applies the element-wise max-pooling operation on the transformed neighbor messages through a fully-connected neural network. 
	The GAT obtains further performance improvement via assigning different learnable weights to neighbor messages.
	\item[-] NCF also obtains competitive recommendation performance, which further validates why the simple collaborative filtering methods can be widely used in recommender systems. 
	On most tasks, our \modela~outperforms the NCF, which demonstrates the graph-structured data are useful for recommender systems.
    \item[-] Our \modela~almost obtains the best performance on multiple datasets. It outperforms the GNN baselines on multi-pair metrics.
    Besides, although the improvement of the \modela~ compared with the GAT is marginal on a few metrics and datasets, the \textbf{Average} values of these metrics of the \modela~are better on all four datasets, which indicates that the \modela~has better generalization performance than the GAT.
\end{itemize}
The essence of recommender systems is to find similarity, and local neighbor nodes often contain such similarity.
Our \modela~aggregates local neighbor messages via high-order feature interactions.
Therefore, the \modela~ can achieve better performance and is more suitable on recommendation tasks.
Overall, these improvements indicate the fact that 
our \modela~can effectively integrate neighbor messages to generate more effective node representations and is more suitable when confronting the graph-structured data. 

\subsubsection{Cross-Domain Recommendation Task}
We also demonstrate the effectiveness of our \modelb~on four pairs cross-domain datasets.
The experimental results evaluated by HR@K and NDCG@K are presented in Table \ref{tab:cross}. 
From these results, we have the following findings.
\begin{itemize}
    \item[-] The collaborative filtering based CD-NCF still obtains competitive recommendation performance via sharing the embedding of overlapped users or items, and it improves the recommendation performance of the CST on all datasets except the TC$\rightarrow$IQI.
    We conjecture that collaborative filtering methods need a lot of data to obtain good performance, while the TC$\rightarrow$IQI have less feedback data.
    It also demonstrates that collaborative filtering is indeed a simple and efficient method in recommender systems.
	\item[-] EMCDR is not an end-to-end method, and the poor performance may result from the accumulation of errors at each step.
	\item[-] EATNN is the state-of-the-art cross-domain recommendation baseline, and it achieves nearly the best results across multiple datasets among these baselines.
	\item[-] By utilizing the graph topology, our \modelb~ improves the recommendation performance compared with various methods.
	It demonstrates that the proposed cross-domain framework combined with the proposed \modela~ is more suitable for the graph-structured data in cross-domain recommendation.
\end{itemize}

\subsubsection{General Cross-Domain Recommendation Task}
Our cross-domain framework is a general framework
that can be applied upon various existing GNN models. 
Here we apply the cross-domain framework
to GCN, GAT, GraphSAGE-mean and GraphSAGE-pooling. 
In order to prove that our cross domain framework is applicable to various GNN models. 
We conduct experiments on 40 tasks ($4\times10=40$, 4 pairs datasets, 10 models). 
The results are shown in Figure \ref{general_framework}. The red lines are the baselines which only use the target training set to train model, also shown in Table \ref{tab:single}, and the blue lines are the cross-domain models which applied the general cross-domain framework.
From the results, we have the following findings:
\begin{itemize}
    \item[-] On most tasks, our cross-domain framework is effective to improve the performance of the single domain models which also demonstrates the cross-domain framework can be applied upon various existing GNN models.
    \item[-] The improvement on GCN is larger than the other four GNN models. The main reason might be that the single domain GCN is significantly weaker than other improved GNN models as showed in Table \ref{tab:single}, so the improvement of other GNN models brought by the cross-domain framework is relatively less than GCN.
    \item[-] The performance of the GraphSAGE-mean and GraphSAGE-pooling is unsatisfying on several datasets, the reason might be that the mean and pooling aggregators are too simple and fewer shared parameters make them difficult to coordinately train in two domains.
\end{itemize}

Overall, we observe that the performance improvement of the cross-domain framework is significant and it is able to improve the performance of base GNN models on different datasets,
which proves that the cross-domain framework is compatible with many GNN models.

\subsection{Ablation Study}
\begin{table*}[!t]
  \centering
  \caption{Results of ablation study on cross-domain recommendation task based on \modelb. ``*'' indicates that the  improvement is statistically significant with the p-value $<$ 0.05 on independent samples t-tests.}
  \resizebox{.7\linewidth}{!}{
    \begin{tabular}{c|ccc|ccc}
    \toprule
    Model & HR@1  & HR@10 & HR@50 & HR@1  & HR@10 & HR@50 \\
    \midrule
          & \multicolumn{3}{c|}{TC$\rightarrow$IQI} & \multicolumn{3}{c}{MO$\rightarrow$MU} \\
    CD-GFM-base & 0.1681 & 0.5914 & 0.9362 & 0.2445 & 0.6989 & 0.9054 \\
    \modelb & \textbf{0.2105*} & \textbf{0.6536*} & \textbf{0.9758*} & \textbf{0.2728*} & \textbf{0.7314*} & \textbf{0.9671*} \\
    \midrule
          & \multicolumn{3}{c|}{ML$\rightarrow$NF} & \multicolumn{3}{c}{MU$\rightarrow$BO} \\
    CD-GFM-base & 0.2178 & 0.6196 & 0.9182 & 0.2756 & 0.6963 & 0.9395 \\
    \modelb & \textbf{0.2243*} & \textbf{0.6247*} & \textbf{0.9228} & \textbf{0.2978*} & \textbf{0.7267*} & \textbf{0.9424} \\
    \bottomrule
    \end{tabular}%
    }
  \label{tab:ablation}%
\end{table*}%
Moreover, for understanding the contribution of the shared node initialization in \modelb.
we construct ablation experiments over \textbf{CD-GFM-base} and \modelb~on four pairs datastes.
\textbf{CD-GFM-base} only uses the domain-specific node representations $\h_{n_s}$ and $\h_{n_t}$ output directly from the \modela~and not to concatenate the initialized input in Equation (\ref{concat1}) and (\ref{concat2}), i.e., 
$\n_s=\h_{n_s},
\n_t=\h_{n_t}.
$
The results are presented in Table \ref{tab:ablation}.
We conduct independent samples t-tests and the p-value $<$ 0.05 indicates
that the improvement of \modelb~over the \textbf{CD-GFM-base} is statistically significant.
The improvement demonstrates that \modelb~model can efficiently take advantage of the domain-shared and domain-specific node representations simultaneously, and obtain the best performance on all datasets, which indicates both two representations matter for the cross-domain recommendation performance.

\section{Conclusion}
In this paper, we first proposed a novel graph neural network model called Graph Factorization Machine (\modela), which utilizes the popular Factorization Machines (FMs) to aggregate multi-order neighbor messages to overcome the shortcomings of the existing GNN models that integrate neighbor messages too simplistic.
Then, we proposed a general cross-domain framework, which can be applied not only to the proposed \modela~to form the cross-domain \modela~(\modelb), but also to other GNN models. 
The extensive experimental results on real-world datasets demonstrate the superior performance of the proposed \modela~model and the general cross domain framework compared with various state-of-the-art baseline methods. 

\ifCLASSOPTIONcompsoc
  \section*{Acknowledgments}
\else
  \section*{Acknowledgment}
\fi

The research work is supported by the National Key Research and Development Program of China under Grant No. 2018YFB1004300, the National Natural Science Foundation of China under Grant No. U1836206, U1811461, 61773361, the Project of Youth Innovation Promotion Association CAS under Grant No. 2017146.

\ifCLASSOPTIONcaptionsoff
  \newpage
\fi


\bibliographystyle{IEEEtran}
\bibliography{IEEEabrv,tkde}
%



%
\begin{IEEEbiography}[{\includegraphics[width=1in,height=1.25in,clip,keepaspectratio]{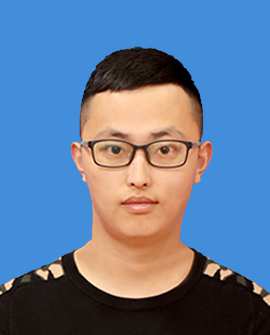}}]{Dongbo Xi}
is currently pursuing his M.S. degree in the Institute of Computing Technology, Chinese Academy of Sciences. He received the B.S. degree in Computer Science and Technology from Nanjing University of Aeronautics and
Astronautics in 2017. He has published some papers in conference proceedings including AAAI, WWW and SIGIR. His research interests include recommender systems and transfer learning.
\end{IEEEbiography}

\begin{IEEEbiography}[{\includegraphics[width=1in,height=1.25in,clip,keepaspectratio]{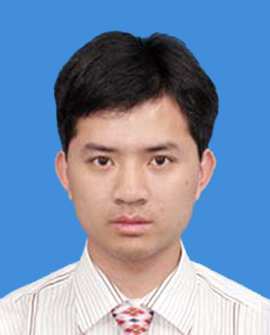}}]{Fuzhen Zhuang}
is an associate professor in the Institute of Computing Technology, Chinese Academy of Sciences, Beijing, China. His research interests include transfer learning, machine
learning, data mining, multi-task learning and recommendation systems. He has published more than 80 papers in some prestigious refereed journals and conference proceedings, such as IEEE Transactions on Knowledge and Data Engineering, IEEE Transactions on Cybernetics, IEEE Transactions on Neural Networks
and Learning Systems, ACM Transactions on Intelligent Systems and Technology, Information Sciences, Neural Networks, SIGKDD, IJCAI, AAAI, WWW, ICDE, ACM CIKM, ACM WSDM, SIAM SDM and IEEE ICDM.
\end{IEEEbiography}

\begin{IEEEbiography}[{\includegraphics[width=1in,height=1.25in,clip,keepaspectratio]{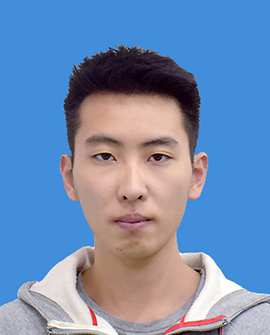}}]{Yongchun Zhu}
is currently pursuing his M.S. degree in the Institute of Computing Technology, Chinese Academy of Sciences, Beijing, China. He has published some papers in journals and conference proceedings including TNNLS, Neural Networks, AAAI, WWW, and PAKDD. He received his B.S. degree from Beijing Normal University, China in 2018. His main research interests include transfer learning, meta learning and recommendation system.
\end{IEEEbiography}

\begin{IEEEbiography}[{\includegraphics[width=1in,height=1.25in,clip,keepaspectratio]{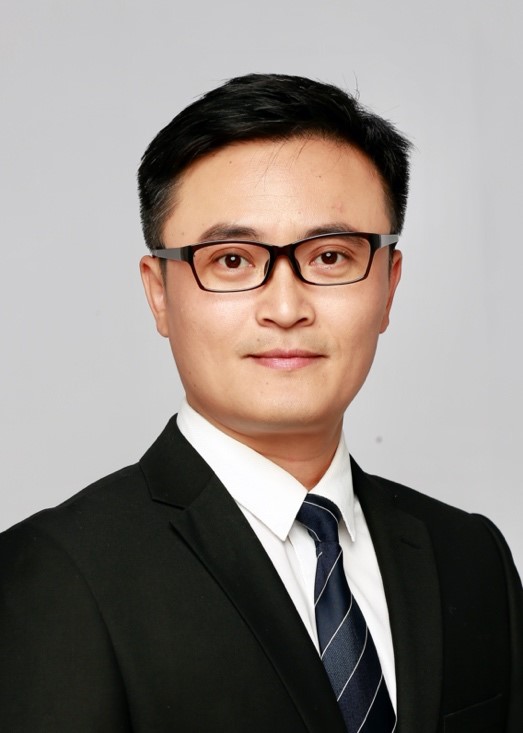}}]{Pengpeng Zhao}
received the Ph.D. degree in computer science from Soochow University in 2008. He is a Professor in the School of Computer Science and Technology at Soochow University. From 2016 to 2017, he was a visiting scholar, working at the Data Mining and Business Analysis Laboratory at Rutgers University. He has published more than 60 papers in prestigious international conferences and journals, including TKDE, ACM MM, AAAI, IJCAI, ICDM, CIKM, DASFAA, and ICME. He was a Program Committee Member of international conferences, such as AAAI, IJCAI, DASFAA, and PAKDD. His current research interests include data mining, deep learning, big data analysis, and recommender systems.
\end{IEEEbiography}

\begin{IEEEbiography}[{\includegraphics[width=1in,height=1.25in,clip,keepaspectratio]{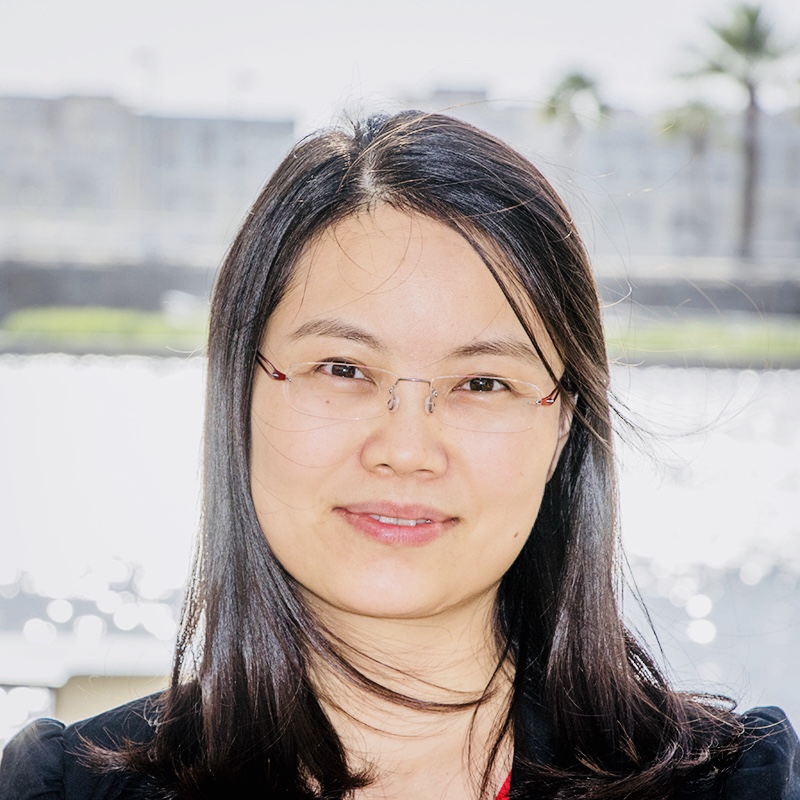}}]{Xiangliang Zhang}
is currently an Associate Professor and directs the Machine Intelligence and Knowledge Engineering (MINE) Laboratory at the Division of Computer, Electrical, and Mathematical Sciences and Engineering, King Abdullah University of Science and Technology (KAUST), Saudi Arabia. She was a European ERCIM Research Fellow at the Norwegian University of Science and Technology, Norway, in 2010. She received the Ph.D. degree in computer science from INRIA-University  Paris-Sud, France, in July 2010. She has authored or co-authored over 130 refereed papers in various journals and conferences. Her current research interests and experiences include machine learning, and data mining.
\end{IEEEbiography}

\begin{IEEEbiography}[{\includegraphics[width=1in,height=1.25in,clip,keepaspectratio]{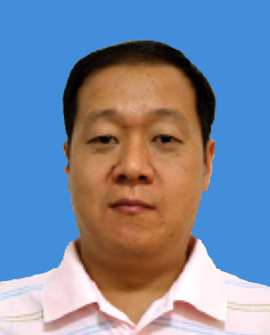}}]{Qing He}
is a professor in the Institute of Computing Technology, Chinese Academy of Sciences, and he is a professor at the Graduate University of Chinese. He received the B.S. degree from Hebei Normal University, Shijiazhuang, China,
in 1985, and the M.S. degree from Zhengzhou University, Zhengzhou, China, in 1987, both in mathematics. He received the Ph.D. degree in 2000 from Beijing Normal University in fuzzy mathematics and artificial intelligence, Beijing, China. Since 1987 to 1997, he has been with Hebei University of Science and Technology. He is currently a doctoral tutor at the Institute of Computing and Technology, Chinese Academy of Sciences. His interests include data mining, machine learning, classification,
and fuzzy clustering.
\end{IEEEbiography}





\end{document}